# Software Licensing in the Cloud Age

Solving the Impact of Cloud Computing on Software Licensing Models

Malcolm McRoberts
Enterprise Architecture Core Technology Center
Harris Corporation GCSD
Melbourne, Florida, USA
mmcrober@harris.com

*Abstract*—Cloud computing represents a major shift in information systems architecture, combining both new deployment models and new business models. Rapid provisioning, elastic scaling, and metered usage are essential characteristics of cloud services, and they require cloud resources with these same characteristics. When cloud services depend on commercial software, the licenses for that software become another resource to be managed by the cloud. This paper examines common licensing models, including open source, and how well they function in a cloud services model. It discusses creative, new, cloud-centric licensing models and how they allow providers to preserve and expand their revenue streams as their partners and customers transition to the cloud. The paper concludes by identifying the next steps to achieve standardized, "cloud-friendly" licensing models.

## I. INTRODUCTION

Software licensing is a major part of IT service delivery. For server software, license fees alone can easily be three or four times the cost of server hardware. In addition, understanding the licensing options offered by vendors and how license terms apply to your operation, and then negotiating the price are all complex and time-consuming tasks. The management of these licenses, including auditing compliance, is a driving factor in IT policy, processes, and architecture.

Cloud computing promises lower costs, improved utilization, increased flexibility and extreme scalability. However, these same features change the software licensing landscape in a disruptive way. Some of these challenges are new and some are magnified by deployment in the cloud.

Organizations that have embraced virtualized infrastructure on a large scale have already experienced the complex and diverse world of licensing on virtual machines. Although based on virtualization, cloud computing represents a more dynamic IT infrastructure and a shift in roles. While it is not possible to capture all the complexity and nuance of this evolving field in a short paper, we can identify key issues that require the attention of the industry as a whole.

### A. Characteristics of Cloud Computing (Benefits)

According to the National Institute of Standards and Technology (NIST) [1], cloud computing is a model for enabling ubiquitous, convenient, on-demand network access to a shared pool of configurable computing resources (e.g., networks, servers, storage, applications, and services) that can be rapidly provisioned and released with minimal management effort or service provider interaction.

This translates to a number of benefits for large and small organizations alike, as well as for individuals. Some of the main drivers for cloud adoption are discussed below:

*1) Over/under provisioning*

Traditional IT resources represent a significant investment and long lead times to acquire and make operational. This not only applies to hardware but also to software. Sizing these resources becomes a dilemma. As shown in Fig. 1, sizing for peak loads results in excess capacity and poor utilization during off-peak periods, while sizing for nominal loads, as shown in Fig.2, will not be able to service peak demands.

By provisioning resources dynamically from large shared pools, cloud services can add or remove capacity on demand and only charge for resources actually used. This makes it possible to automatically scale resources (within minutes) to match the actual demand, as shown in Fig. 3. This type of auto-scaling requires fully automated provisioning, including any licensed software required.

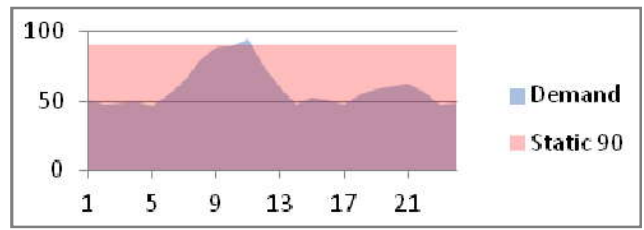

Figure 1. Over-Provisioned Resources for Peak Demand

395

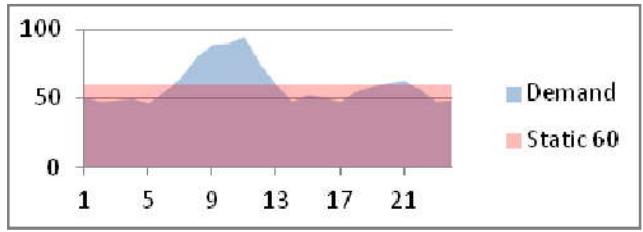

Figure 2. Resources Provisioned for Nominal Demand

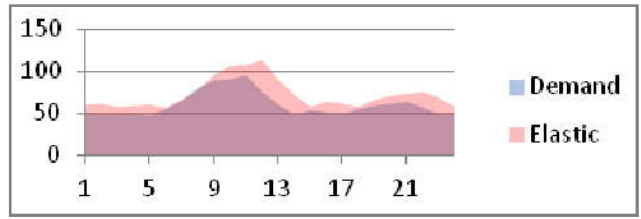

Figure 3. Cloud Resource Auto-Scaling

*2) Agility*

Cloud computing can greatly enhance the ability of small companies and small projects within large companies to develop and deploy new applications quickly and cheaply. This agility comes from the ability of projects or individual developers to rapidly provision and control the resources needed to develop, test, integrate, and deploy their applications. Control typically includes network configuration, virtual machine life cycle, and even root access to guest operating systems. Public or community clouds are ideal for small companies; larger companies may choose to provide cloud services internally.

Virtualized infrastructure is often equated with cloud computing by corporate IT leadership. Through the use of virtualization, mature IT organizations can often provide relatively rapid (hours/days) provisioning of development resources, whereas, in a cloud, users can provision their own resources in minutes via APIs and web portals. Typically, when IT is responsible for provisioning development resources, it retains control of the resources providing developers with the limited access, in contrast to the delegated control provided by a true cloud. Central management of these resources becomes problematic, and most organizations end up with large virtual machine junkyards.

The benefits of prototyping in the cloud are so great that IT organizations must either embrace this new approach or be bypassed by shadow IT in the public cloud (software licensing issues will be addressed in subsequent sections).

*3) Budgeting for software costs*

A traditional IT system typically requires a large, up-front capital investment (CapEx) that is amortized over several years and does not provide value until the system is operating at capacity. Cloud computing provides the ability to avoid purchasing infrastructure and instead pay for the cloud service as an operational expense (OpEx). As a result, a business conserves capital much like it would by leasing its office space. In addition, cloud expenses grow in proportion to the capacity (load) of the system.

The downside of this is that the scale is not fixed up front, and cloud charges can grow more rapidly than expected, a problem if not tied to a proportionally growing revenue stream.

*B. Cloud Threats to Commercial Software*

Cloud computing represents a more dynamic and flexible approach to providing IT services. It involves new distributed architectures, new ownership and control models, and new acquisition and pricing mechanisms. The disruptive influence of cloud computing on software licensing as been discussed in [2] and [3].

Traditional licensing models for commercial software products often make it difficult or impossible to use these products in the cloud. Independent software vendors (ISVs) that fail to adopt more flexible licensing schemes risk losing customers to other vendors. The more significant risk for vendors is that customers will move to completely open source products and avoid the issue entirely, as the most successful Internet companies have already done.

The flip side is that organizations struggling to adopt cloud computing or even large-scale virtualization are finding that software licensing is holding back their efforts. Some of the problems encountered are discussed below:

*1) On-demand provisioning*

Cloud computing allows users to rapidly provision servers via APIs or GUI tools. Prior to cloud computing, users had to request resources through their IT organization. This allowed IT to ensure that software was properly licensed, and to perform server-specific activation. In the cloud, software licenses must either be automatically provisioned and managed by the cloud, or subscribers must provide their own, often a manual process.

*2) Elastic scaling*

Clouds provide support for systems that automatically scale up when demand increases. Many products support the notion of a network pool of licenses, which is normally based on purchasing a fixed capacity (see Fig.1 and Fig. 2). In cloud bursting, scaling may involve more than one cloud infrastructure provider, further complicating license management.

*3) Short-term resource (license) allocation*

Clouds can support both long-running, static deployment as well as short-term, task-based workloads. Resources can be provisioned, used, and then released all in the span of a few minutes. Most licensing schemes are not designed to handle this type of usage pattern.

*4) Shared resource pools (multi-tenancy), delegation*

The pricing of many corporate software agreements is based on the organizational affiliation of the software user. Public and community clouds service many tenants (subscribers) from large, shared pools of resources. A cloud provider can only supply licenses if the license allows delegation to their tenants.

5) Compliance, indemnification and shadow IT

Software license compliance, along with security compliance, is a key factor driving centrally managed IT. While some software products provide their own license enforcement mechanisms, it is ultimately the responsibility of the user's company to ensure license compliance and maintain auditable compliance records. License agreements typically include the right of the vendor or representative to conduct audits and levy fines. SIIA [4] and BSA [5] are leading organizations that promote the active policing of license compliance. Large fines and even jail time for executives can result from infringement.

While cloud computing promises to allow users to provision and control their own resources, releasing them from the constraints of their IT group, it may also bypass IT's role in managing license compliance (and cyber security).

Even shops with centrally managed virtual infrastructure find that if users can install software themselves, their corporate license footprint can rapidly outpace their license agreement. Getting internal users to release licenses they no longer need is an ongoing challenge, as is getting them to release virtual machines (VMs).

A bigger threat to compliance comes from shadow IT. Projects that perceive corporate IT as too restrictive, expensive, slow, etc., often resort to obtaining resources outside corporate governance. In the past, this involved building labs with small hardware purchases, and using unauthorized network connections. Now anyone can create—in minutes—shadow IT in the cloud which is not tracked by corporate IT. Since most companies have no sanctioned mechanism for procuring cloud services and most cloud providers have a bring-your-own-license (BYOL) policy, users and their companies may be put at risk.

6) License enforcement

Many products provide their own license enforcement mechanism; however, most of these are problematic for use in the cloud. These mechanisms may depend on features difficult to provide in the cloud, such as hardware keys (dongles), physical server IDs, CPU class, and global user identity. Where schemes can be implemented on one cloud, they cannot generally span multiple clouds (hybrid model).

7) Adoption-led acquisition models

Traditional software licensing has evolved alongside traditional acquisition models. In this model, trades are performed and products are selected and purchased early in the development of a system. Changing out products can result in substantial loss of investment and impact schedules, and is only considered in extreme cases.

Many organizations are now opting for an adoption-led approach to acquisitions [6] and expect to license on a try-before-you-buy basis. For instance, a company might deploy several different web meeting tools and license one or more depending on usage levels after a trial period.

Time-limited evaluation licenses attempt to address this need, but still have a large-scale, traditional license as the end goal.

8) Open source

Software licensed under open source models as defined by the Open Source Initiative [7] is in widespread use throughout the industry and continues to grow. This includes large shares in areas such as operation systems, databases, application servers, and hypervisors. Many factors drive the adoption of open source, including the cost and complexity of commercial licenses, the proliferation of third-party companies offering training and support services, and freedom from vendor lock-in.

Market leaders in cloud computing and Internet-based services in general have constructed their offerings almost exclusively from a combination of their own software and open source offerings. Open source allows these companies to start small and grow fast.

The shift to cloud computing only serves to highlight the advantages of open source licensing.

9) Commodity pricing

Pricing and terms for commercial software are typically negotiated between an ISV and the individual licensee, based on each party's negotiating leverage. The trivial reproduction costs for software allow vendor account representatives tremendous freedom to set terms, using special pricing to lock in large accounts and effectively block use of competing products while integrating their products into the customer's operations and encouraging product-specific investment in training, custom tools, etc.

Larger customers are able to negotiate better pricing, reduce costs, and increase their competitiveness. Favorable reseller agreements directly impact the cost of the reseller's own products. Long-term, enterprise-wide standardization on a product enables corporate IT to develop streamlined support processes and avoid disruption in service.

The natural view of software in the cloud is as just another resource, which is billed at a standard hourly rate like CPU or storage. This threatens to undermine the power base of both the vendor and customer and introduces a degree of separation between them.

*C. Rights versus Revenue*

Software development requires a substantial investment of both time and money. Copyright laws are designed to encourage investment by giving creators the legal right to profit from their creation. Once a product has been developed, the focus should be on maximizing revenue generation. Restrictive license agreements and rigid license enforcement mechanisms may do a good job of protecting a vendor's intellectual property rights, but they also may negatively impact sales. Dongles are among the mechanisms that have been rejected by software customers.

Open source software licenses are designed to allow the software to be used by anyone free of charge, but most open source software projects are sponsored by commercial firms.

These firms have successful business models that are not dependent on the enforcement of license agreements.

A license scheme, pricing model, and enforcement mechanism should be designed to maximize revenue, not exclusively to protect property rights.

## II. CLOUD SERVICE MODELS AND ROLES

Cloud computing is a rapidly evolving space that does not fit cleanly into a simple model. A wide variety of physical deployments, network topologies, business models, and software architectures are possible, including those yet to be conceived. Much like mobile phone service requires a complex network of service provides and vendors, cloud services are also made up of many parts. For cloud computing to mature, it must achieve the same level of industry standardization and consumer simplicity as the mobile phone industry.

### A. NIST Models

NIST has led the way in developing a standard taxonomy for cloud computing [1], defining essential characteristics, service models, and deployment models. The NIST service model consists of three layers: infrastructure, platform, and software. All three layers are involved in delivering value to customers, but all may not take the form of a service. Each service layer may involve licensed software, thereby incurring license charges.

1) Infrastructure as a Service (IaaS)

IaaS consists of CPU, storage, networking, and operating systems and supporting services provisioned on demand and delivered via the web. IaaS typically uses a hypervisor to provide virtual machine instances.

2) Platform as a Service (PaaS)

PaaS is the least understood model. The platform consists of frameworks and services that host or support application software, but not the user-facing applications themselves. Databases, application servers (middleware), and business service APIs are common part of a PaaS.

3) Software as a Service (SaaS)

In this model, users run applications and consume content over the network (Internet or intranet), much as they would use software installed on their personal devices. Some of the largest consumer content providers host their services on public cloud infrastructure.

### B. Roles

The service layers previously described can all be provided by a single organization, or services can be layered on top of services from different providers, creating a type of supply chain or service map. Each step in the chain is defined by a service agreement that may involve software licensing. This perspective differs from the NIST taxonomy, which is focused primarily on security and systems engineering. Roles include:

1) Subscriber

A subscriber is the entity that contracts for cloud services. These services may be consumed by the subscriber's IT systems or be directly used by its staff, partners, or customers.

2) End user

An end user is the entity who uses software or content delivered over the network. Cloud services ultimately derive value by delivering service to end users.

3) ISV

An ISV is a firm that develops proprietary software with the intent to license the software at a profit. ISVs typically have a portfolio of related products and supporting services, such as training and integration.

4) SaaS provider

The SaaS provider delivers application and content to end users over the network.

5) PaaS provider

The PaaS provider hosts applications developed by others and provides tools and frameworks for use in building applications.

6) IaaS provider

The IaaS provider delivers servers (usually virtual) with operating systems, storage, and network connectivity on demand over the Internet.

7) Identity provider

An identity provider allows users to establish a network-wide identity and access resources from multiple service providers based on a single authentication. Identity is used to manage access to personal data, billing, and user-based software/content licensing.

### C. Relationships

The aforementioned roles are common to most cloud services, but the relationships between them determine software ownership and licensing patterns.

NIST defines four deployment models: public, community, private, and hybrid. Rather than defining these strictly in terms of network topology, we define them in terms of the relationship between the service provider and the subscriber. In a public cloud, the service is vended as a commodity on the open market. Any customer (subject to legal boundaries) can subscribe for the service. Public cloud providers may license their own proprietary software as part of their service in a pay-as-you-go (PAYG) model. The provider may also act as an ISV, licensing the same software outside of their cloud. Public cloud providers can supply licensing for third-party software or require subscribers to BYOL. In any case, subscribers are ultimately responsible for their own license compliance.

In a community cloud, the subscribers come from a community based on existing relationships with the provider and other subscribers. Such an organization may subsidize the cloud service, negotiate communitywide licenses, or even sponsor software development. They may also operate private networks to provide secure communication channels within the community.

In a private cloud, the provider and subscriber belong to the same organization, and services are typically only provided in the organization's intranet. In most cases, subscribers are not charged for service, and often usage is not even metered. Private clouds are usually operated by corporate IT, who is also accountable for its company's license compliance.

*D. Portablity – Hybrid Cloud*

Hybrid clouds combine private and public cloud services. The hybrid cloud can handle workload spikes by augmenting private cloud resources with resources from the larger public cloud. Cloud industry standards bodies have been working on the problem of workload portability between clouds, and commercial tools and services are available to manage hybrid clouds. But workload portability does not guarantee license portability. In practice, license terms and management tools differ between private and public clouds.

### III. TRADITIONAL SOFTWARE LICENSE MODELS

This section discusses the most common types of commercial software licenses and issues arising from applying these in a cloud-based environment. This topic has also been discussed in [14].

*A. Host ID-Based Model*

In this scheme, one or more server hardware components are queried to generate a unique host ID. The customer obtains a license key from the vendor that is tied to the host ID. This scheme does not work well for cloud-based systems for several reasons. Most clouds are based on virtualization, meaning that the hardware is abstracted (hidden) from the cloud tenants. Even if a host ID can be generated for the physical device, the virtual server may be migrated to a different hardware as a function of cloud management.

Clouds are also designed to support automated provisioning and elastic scaling, which are not possible if the software vendor must be contacted each time a host is provisioned or migrated.

*B. Token-Based Model*

In this scheme, a physical license key, such as a dongle or copy-protected CD, must be physically attached to the host running the software. The advantage for workstation users is that the key can be easily moved between machines. Since clouds and most virtualized infrastructures do not allow client physical access to servers, this kind of scheme is generally not usable.

*C. Named User Model*

In this scheme, a license is tied to a specific user. That user is licensed to use the software on any device and, in some cases, multiple devices at the same time. Determining if a user is licensed for the software requires that the user authenticate against an identity provider in the domain where the software will run. Traditionally, licenses are tied to a company's internal directory service, but cloud-based services may utilize Internet identity providers. Services that deliver music and videos typically use this type of license. Managing identity-based licenses across domains, such as in a hybrid cloud, may be difficult.

Named user licenses are often sold to a company as a pool from which they can then assign and transfer as needed.

*D. Concurrent Users Model*

In this scheme, the customer purchases a pool of licenses, which users check out as they would (concurrent) copies of a library book. The definition of concurrency can depend on the type of software involved. A compiler, a word processor, and an application server may have very different concepts of concurrent use.

License check-in/checkout is usually automatic, tied to a web session or the start and stop of an application. This is usually implemented via application hooks that call out to a license manager service. Unfortunately, most software provides its own license manager, making consolidated management difficult. This type of licensing typically works the same in the cloud as in a private network, provided that the license terms permit this. In multi-tenant cloud services, it may be desirable to subdivide the license pool to enforce quotas on tenants. This may not be supported by license manager software. Among traditional license schemes, this is probably the best fits for the cloud computing paradigm.

*E. Hardware Capacity-Based Model*

The amount of work (e.g., transactions) that a software system can perform in a given time is bounded by the hardware on which it runs. This type of license equates software value with computing power. The formulas used to calculate power for licensing purposes are somewhat arbitrary and vary from vendor to vendor. This type of license does not usually provide any automated enforcement.

Optimum use of this type of license requires that hardware be scaled to fit expected load and dedicated for use by a single software package. Dedicating physical hardware to a specific application is contrary to the principle of cloud computing.

Fortunately, many vendors provide formulas for virtual machines and specific cloud services. IBM's bring your own software and license (BYOSL) strategy assigns power units to each type of Amazon EC2 instance. While this type of license supports use in the cloud, the allocation of licenses is still usually static.

*F. Site-Wide Model*

Site licenses are negotiated based on the general size of the customer site. The premise is that the site corresponds to a local or campus area network and the software can be freely installed and used on this network. Cloud computing explicitly hides physical site and network boundaries making this type of license a bad fit for cloud computing in general. Even private clouds usually support users from any location as long as they are on the private network, making this equivalent to an enterprise-wide license.

*G. Enterprisewide Model*

In this model, an ISV licenses its software to an entire enterprise. This type of license typically covers installation and use of software by employees of the enterprise on equipment owned by the enterprise. Exceptions may include employee-owned devices, known as bring your own device (BYOD) [9], use by subcontractors and employee home use.

Enterprise licenses are not intended to be used by service providers to resell the software to their customers. Some vendors have recognized this need and have provided licenses specifically for service providers [10].

While cloud providers may offer pay as you go licensing, large customers that have favorable enterprise license terms may prefer to use their own license. Some cloud providers and ISVs have partnered to provide explicit support for BYOL [11][12].

*H. "Commercial" Open Source Model*

Open source software [7] may be copied and used by anyone free of charge. In spite of this, there is a large market for firms providing open source software. These firms do not sell licenses to software that is already free, but they do provide services that enterprise customers find valuable including:

1) Packaging, certification, accreditation

Vendors providing these services take snapshots of multiple open source baselines, integrate them, test them, and certify them. In some cases they may also obtain security accreditations needed for use in government systems.

2) Indemnification

Complex software products often embed components from multiple sources. The vendor evaluates all the licenses that apply to the integrated product and "indemnifies" the customer from any legal liability from embedded products.

3) Security patches (subscriptions)

The security posture of any IT systems depends on the timely identification and installation of critical security patches. Vendors provide subscription services to ensure that systems are up to date.

4) Support

Some vendors provide Tier 3 technical support, training, and consulting service, just like a large ISV. In the case of open source software, there may be multiple companies competing to provide support for the same software.

Commercial open source is sold under terms similar to those for commercial software. Since the price of this software tends to increase with the level of use, it may become more cost effective for an organization to provide these services internally. This would generally apply to service providers, although subscribers may choose to purchase software in this way to use in the cloud.

*I. Free Open Source Model*

Rather than pay commercial providers for open source, companies can obtain the software in source or binary form directly from open source project sites. This type of licensing is ideal for cloud service providers as well as subscribers. There are no restrictions on who may use the software, number of users, location, or size of hardware. Adopters of the free open source model take on the added responsibilities of tracking and integrating new releases, training and maintaining knowledgeable support staff, and ensuring the quality and security of the software they are using. Community support forums take the place of Tier 3 product support, but can often be equally responsive and helpful.

Leading internet companies have embraced open source both as users and as contributors. In order to prosper, commercial ISVs must provide unique features and exceptional quality to justify their cost, while also adopting licensing models that scale up and down for use in the cloud.

*J. Ownership – Copyright Holder Model*

Leading cloud service providers typically use a mixture o open source software and software from in-house development projects. Companies that hold the software copyrights either through creation or acquisition can use the software in any way they choose. Also ISVs may choose to become service providers to give their customers the option to pay for their software in the form of a service.

IV. CLOUD-CENTRIC LICENSE MODELS

While the issues described earlier still pervade the industry, this section addressed cloud-centric licensing models currently in use by leading cloud providers.

*A. Amazon Web Services – DevPay*

Amazon has always been and continues to be a pioneer in the cloud computing market space. This extends to Amazon's innovative billing service, DevPay [13]. DevPay allows software vendors of any size to deliver their software as a service within the Amazon cloud and use DevPay to manage access and billing for that service. Users of DevPay-enabled software must have an Amazon account and are billed for their software use by Amazon, much as mobile providers bill for applications. Software charges are normally computed as a surcharge on Amazon resources used, including virtual machine instances, storage, and bandwidth. Users can view rates and track charges on Amazon's dashboard. Software providers may also restrict access to registered customers. This mechanism is enabled by adding a few simple web service calls into the application software. This is especially attractive for small developers who can generate revenue, while avoiding the complexity of licensing and billing.

Amazon uses this same mechanism to bill for some of their higher value, platform services, which are built on top of the same infrastructure resources.

*B. IBM SmartCloud*

IBM Smart Cloud is a service based on hourly charges for virtual machine instances. IBM has chosen to lease a variety of products from their own middleware portfolio on a pay-as-you-go model. Machine images with these products and their hourly rates are listed in the Smart Cloud catalog. In addition, to encourage development in the Smart Cloud, development use only (DOU) licenses are available for select products free of charge.

*C. Microsoft Azure*

The Microsoft Azure cloud provides a platform for deploying custom applications as well as a host of services based largely on Microsoft products that have been engineered to work in the Azure cloud. Most of Microsoft's software products are available as services on Azure.

*D. iTunes - Apple ID*

Apple is a pioneer in end-user computing devices ranging from Macs to iPhones. The appeal of these devices is largely driven by their integration with Apple's cloud service. Apple has also pioneered the idea of software and content that is not bound to a device, but rather to a user identity in the cloud. For Apple, this is the user's Apple ID, which provides access to content and software belonging to that user from any compatible device. This is similar to intranet, named user licenses, but is global in scope, since it is based on a global ID and an Internet-based cloud service.

V. STEPS TOWARD STANDARDIZING CLOUD-FRIENDLY LICENSING

Cloud computing is revolutionizing the way organizations pay for and use their IT resources. This paper has shown that while cloud computing has the potential to simplify the licensing and use of software, it has, in fact, only added to the problem.

For commercial software vendors to successfully move into the cloud age, they must now work as a group with cloud providers to standardize licensing in the cloud. Standards-developing organizations (SDOs) should govern the activity. A successful solution must address legal and financial concerns, as well a technical aspects of software licensing in the cloud. To do this effectively, the solution should provide the following:

*A. Consistency and Confidence*

Customers need to be able to use licensed software with confidence that their use will comply with the license; that the costs are understood, fair, and predictable; and that licensing is consistent between products and between cloud providers. Customers should be able to use software without resorting to guesswork or lawyers.

*B. Inter-Cloud Portablity*

Cloud industry standards bodies have been working toward standards to allow data and workloads to move seamlessly between clouds, both public and private. Software licenses must migrate seamlessly as well.

*C. Protection of Intellectual Property*

Cloud license agreements and technical standards must provide ISVs with the mechanisms they require to track the use of their IP, audit license compliance, collect fees and prevent software piracy.

*D. License Model Options*

ISVs must be able to choose from a range of standardized license models, the one that best fits their product. These would include iTunes and DevPay style licenses.

*E. Pricing Flexibility*

Service providers and subscribers should have the option to negotiate software pricing with ISVs or agents, or to pay list. Service providers that provide licenses should give subscribers the option to use their own.

*F. Ability of Service Providers to Perform as Agents*

ISVs must be able to delegate license management (tracking and enforcement), accounting and billing to service providers, based on standardized mechanisms. These mechanisms should support pass through to other service providers.

*G. Automation through Standard APIs*

Service providers need standard APIs to automate license management and make license management an integral part of their resource management systems. ISVs require standard APIs to allow use of their licensed software to be transparently managed by any service provider. APIs are also needed to support pass through and migration between service providers.

Once the industry has defined what makes up a "cloud-friendly" software license, the final step is to provide endorsements, much like OSI does for open source licenses. This endorsement would allow providers and consumers alike to select and use products based on functionality and price, without concern for the subtleties of license terms.

REFERENCES

[1] National Institute of Standards and Technology (NIST), Special Publication 800-145, "The NIST Definition of Cloud Computing" URL: http://csrc.nist.gov/publications/nistpubs/800-146/sp800-146.pdf

[2] Leah Gabriel Nurik, "Cloud Computing Disrupting Software Licensing, Pricing", Channel Insider, 2010-10-22 http://www.channelinsider.com/c/a/Cloud-Computing/Cloud-Computing-Disrupting-Software-Licensing-Pricing-557629/

[3] Adam Stone, "IN MOVING TO THE CLOUD, SOFTWARE LICENSING PROVES A STUMBLING BLOCK", NextGov, February 28, 2011 URL:http://www.nextgov.com/cloud-computing/2011/02/in-moving-to-the-cloud-software-licensing-proves-a-stumbling-block/48596/

[4] Software and Information Industry Association URL: http://siia.net/

[5] Business Software Alliance, "Software Piracy and the Law" URL:http://www.bsa.org/country/Anti-Piracy/Know%20the%20Law.aspx

[6] Stuart Charlton, "Software Licensing In The Cloud", CloudWorld 2009, Aug 13, 2009
URL: http://www.slideshare.net/slideshow/embed_code/1857667

[7] Open Source Initiative, "The Open Source Definition"
URL: http://opensource.org/osd

[8] Joaquin Gamboa and Marc Lindsey, "Cloud-enabling your software licenses", ComputerWorld, September 8, 2008
URL: http://www.computerworld.com/s/article/323517/Cloud_enabling_Your_Software_Licenses

[9] PETER SILVA, "Will BYOL (Bring Your Own License) Cripple BYOD?", Virtualizaton Journal, JULY 4, 2012
URL: http://virtualization.sys-con.com/node/2305543

[10] Microsoft Service Provider License Agreement
URL: http://www.microsoft.com/hosting/en/in/licensing/splabenefits.aspx

[11] Amazon Relational Database Service Pricing
URL: http://aws.amazon.com/rds/pricing/

[12] IBM Licensing for Amazon Cloud: (BYOSL)
URL: http://www-01.ibm.com/software/lotus/passportadvantage/pvu_for_Amazon_Elastic_compute_cloud.html

[13] Amazon DevPay
URL: http://aws.amazon.com/devpay/

[14] Dave Roberts, GIGAOM, Jul 8, 2012 http://gigaom.com/cloud/dont-let-your-cloud-app-become-a-software-licensing-hostage/

[15] http://portal.bsa.org/globalpiracy2011/